\documentclass[a4paper,11pt]{article}
\usepackage{graphicx,amssymb,bm,latexsym}
\makeatletter
\@addtoreset{equation}{section}

\makeatother

\newcommand{\bea}{\begin{eqnarray}}
\newcommand{\eea}{\end{eqnarray}}
\newcommand{\vs}[1]{\vspace{#1 mm}}
\newcommand{\hs}[1]{\hspace{#1 mm}}
\renewcommand{\a}{\alpha}
\renewcommand{\b}{\beta}

\newcommand{\G}{\Gamma}
\renewcommand{\d}{\delta}

\newcommand{\s}{\sigma}
\renewcommand{\t}{\theta}
\newcommand{\vp}{\varphi}
\newcommand{\la}{\lambda}
\newcommand{\pa}{\partial}
\newcommand{\nn}{\nonumber\\}
\newcommand{\p}[1]{(\ref{#1})}

\newcommand{\ta}{\tilde\a}
\newcommand{\tb}{\tilde\b}
\newcommand{\tL}{\tilde\Lambda}

\begin{document}

\begin{titlepage}

\begin{flushright}
KU-TP 056 \\
\end{flushright}

\vs{10}
\begin{center}
{\Large\bf Beta Function and Asymptotic Safety in Three-dimensional Higher Derivative Gravity}
\vs{15}

{\large
Nobuyoshi Ohta\footnote{e-mail address: ohtan@phys.kindai.ac.jp}} \\
\vs{10}
{\em Department of Physics, Kinki University,
Higashi-Osaka, Osaka 577-8502, Japan}

\vs{15}
{\bf Abstract}
\end{center}

We study the quantum properties of the three-dimensional higher derivative gravity.
In particular, we calculate the running of the gravitational and cosmological constants.
The flow of these couplings shows that there exist both Gaussian and nontrivial
fixed points in the theory, thus confirming that the theory is asymptotically safe.
It is shown that the new massive gravity or $f(R)$ gravity in three dimensions
do not correspond to the fixed point within the approximation that the coefficients
of the higher curvature terms are not subject to the flow.
The fixed point value of the cosmological constant is found to be gauge-independent,
positive and small.
We also find that if we start with Einstein term with negative sign,
the fixed point only exists when the coefficient of the Einstein term has positive sign.

\end{titlepage}
\newpage
\setcounter{page}{2}

\section{Introduction}

One of the long standing important problems in theoretical physics is to understand
quantum properties of gravity. The usual Einstein gravity is known
to be non-renormalizable in four and higher dimensions. If we add
Ricci and scalar curvature squared terms, the theory becomes renormalizable,
but then the unitarity is lost~\cite{Stelle}. So this higher derivative gravity
does not appear to make sense as a physical theory.

Recently a very interesting proposal has been made that the addition of such higher
order terms to three-dimensional gravity makes the theory unitary if the coefficients
are chosen appropriately~\cite{BHT1} (with ``wrong sign'' Einstein term).
The usual Einstein gravity does not have any propagating mode, but the addition of
these terms introduces propagating massive graviton around flat Minkowski and
curved maximally symmetric spacetimes [anti-de Sitter (AdS) and de Sitter (dS) spacetimes].
Another theory of massive graviton with Lorentz-Chern-Simons (LCS) term has
long been known as topologically massive theory~\cite{Top}, which breaks the parity.
The above new theory is an ordinary parity-preserving theory and is called new massive
gravity. This is very interesting in that we have really dynamical theory of gravity
that is unitary even though higher derivative terms are included.
Since then, various aspects of the theory have been investigated.
Linearized excitations in the field equations were studied in~\cite{LS}.
Unitarity is proven for Minkowski spacetime in~\cite{Oda,Deser,GST}, whereas
it is discussed in \cite{BHT2} for maximally symmetric spacetimes.
A complete classification of the unitary theory for the most general action with arbitrary
coefficients of all possible terms is given in \cite{Ohta1}.
The partial result of unitarity condition on the flat Minkowski spacetime was known
for the usual sign of the Einstein theory~\cite{NR}.
Unfortunately the new massive gravity turned out to be non-renormalizable
though the general theory with arbitrary coefficients for the quadratic curvature
terms is renormalizable~\cite{Deser,BHTE,Ohta2}.

However, even if the theory is not renormalizable, it is possible that
a theory has a nice ultraviolet property if it has an ultraviolet fixed point.
A non-renormalizable theory may be made effectively renormalizable by a rearrangement
of the perturbation series or by addition of higher derivative terms.
However terms of finite order in the perturbation series then contain what appear
to be unphysical singularities. Such unphysical singularities may be almost certainly
avoided if the couplings approach a fixed point in the ultraviolet energy.
This property is known as asymptotic safety~\cite{Weinberg},
and it seems that this is the only way to make sense of gravity theory known to date.
In particular the asymptotic safety is a wider notion than the renormalizability.
Any theory will always have a fixed point at the origin.
If this is the only suitable fixed point with ultraviolet critical surface of
nonzero dimensionality, then the asymptotic safety requires that the couplings
lie on this surface. In order for the trajectory of the coupling constants
to hit the origin in the high energy, all couplings with negative dimensionality
(in powers of mass) should vanish. These are precisely the non-renormalizable interactions,
so this sort of theory must be renormalizable in the usual sense.

It should be very interesting to examine what quantum properties the above theory
of higher derivative gravity has since it is dynamical theory of gravity and preserves
parity as the ordinary gravity theories.
The renormalization group (RG) properties of four-dimensional gravity are studied,
for instance, in \cite{Reuter,CPR,BMS2}.
Those of the topological gravity in three dimensions have been studied in \cite{PS},
but as far as we are aware, these have not been examined for general higher derivative
theory in three dimensions related to the above new massive gravity.
Discussions of four-dimensional higher curvature theories are given in
Refs.~\cite{BGO,LR,CP,BMS1}.
A simple RG flow argument is used in the case of Horava gravity in~\cite{GM},
and also 3D $f(T)$ gravity is analyzed in \cite{GSV}.
In this paper we would like to examine if the three-dimensional theory is asymptotically safe,
and the theory at the fixed point has any relation with the above new massive gravity.

In the next section, we briefly summarize the Wilsonian RG approach
to gravitational theory. Here we have to introduce the cutoff to regularize the theory,
and must be careful about what is meant by the cutoff since the definition of the cutoff
in general uses a metric which is dynamical. We just follow the standard way to
use background method.
In sect.~\ref{secqe}, we introduce the gauge fixing and the corresponding Faddeev-Popov (FP)
terms, and derive the quadratic part of the action in the fluctuations.
In sect.~\ref{seceval}, we then evaluate the resulting functional traces and derive
one-loop beta functions.
In sect.~\ref{secfix}, we discuss what flows and fixed points we obtain. In our analysis,
we restrict ourselves to the flow in the gravitational and cosmological constants
since it appears that the coefficients of the higher derivative terms can be set to
constant, and then the analysis becomes simple. This analysis is done for the usual
and opposite signs of the Einstein term. In both cases, we find that there are
Gaussian fixed point at the origin and nontrivial fixed points as the ultraviolet
attractive points.
Even without the beta functions for the higher curvature terms,
our analysis indicates that the new massive gravity or a special case of $f(R)$ gravity
in three dimensions do not correspond to the fixed point.
We also find that if we start with Einstein term with negative sign,
the fixed point only exists when the coefficient of the Einstein term has positive sign.
The final section~\ref{secconc} is devoted to discussions and conclusion.

\section{Wilsonian method for renormalization group}
\label{secrg}

In the Wilsonian RG, we consider that the effective action
describing physical phenomenon at a momentum scale $k$ can be thought of
as the result of integrating out all fluctuations of the fields with momenta
larger than $k$. We can regard $k$ as the lower limit of the functional integration
and call it the infrared cutoff. The dependence of the effective action on $k$
gives the Wilsonian RG flow.

There are several ways of implementing this. We follow Ref.~\cite{PS} and
suppress the contribution of the field modes with momenta lower than $k$
by modifying the low momentum end of the propagator and leaving all
the interactions unaffected.

We start with a bare action $S[\phi]$ for some fields $\phi$, and add a suppression
term $\Delta S_k[\phi]$ which is quadratic in the field.
We choose some differential operator ${\cal O}$ whose eigenfunctions $\vp_n$,
defined by ${\cal O}\vp_n=\la_n \vp_n$, are taken as a basis in the functional
space we integrate over:
\bea
\phi(x) = \sum_n \tilde\phi_n \vp_n(x).
\eea
The additional term is written as
\bea
\Delta S_k[\phi] = \frac12 \int dx \phi(x) R_k({\cal O})\phi(x)
= \frac12 \sum_n \tilde\phi_n^2 R_k(\la_n).
\eea
The kernel $R_k(z)$ is introduced as a cutoff, and is chosen to be
a monotonically decreasing function in both $z$ and $k$, namely
$R_k(z) \to 0$ for $z \gg k$ and $R_k(z) \neq 0$ for $z \ll k$.
Define a $k$-dependent generating functional of the connected Green functions by
\bea
e^{-W_k[J]} = \int {\cal D}\phi \exp\Big[ -S[\phi]-\Delta S_k[\phi]-\int dx J\phi\Big],
\eea
and a modified $k$-dependent Legendre transform
\bea
\G_k[\phi] = W_k[J]-\int dx J\phi-\Delta S_k[\phi].
\eea
In the limit of $k\to \infty$, this functional tends to the usual
effective action $\G[\phi]$.

We assume that $\G_k$ admits a derivative expansion
\bea
\G_k[\phi, g_i] = \sum_{n=0}^\infty \sum_i g_i^{(n)} {\cal O}_i^{(n)}(\phi),
\eea
where $g_i^{(n)}$ are coupling constants and ${\cal O}_i^{(n)}$ all possible
operators constructed with the field $\phi$ and $n$ derivatives compatible
with the symmetries of the theory.
The index $i$ is used to label different operators with the same number of derivatives.
At the one loop, $\G_k$ is given by
\bea
\G_k^{(1)} = S+\frac12 \mbox{Tr} \log(S^{(2)}+R_k),
\eea
where $S^{(2)}$ denotes the second variation of the bare action.
We then obtain
\bea
k\frac{d\G_k^{(1)}}{dk} = \frac12 \mbox{Tr}(S^{(2)}+R_k)^{-1} k\frac{dR_k}{dk}.
\eea
The factor $k\frac{dR_k}{dk}$ goes to zero for $z>k^2$.
One can obtain the one-loop beta functions from this functional equation.

We apply this method to higher derivative gravity in three dimensions.
Here one has to be careful about what one means by cutoff because the definition of
a cutoff generally makes use of a metric which in turn is to be treated as a dynamical field.
We follow \cite{Reuter, PS} and use the background method, and effectively
replace the dynamical metric $g_{\mu\nu}$ by a spin-two field $h_{\mu\nu}$ propagating
in a fixed background $\bar g_{\mu\nu}$.
The background metric can be used to unambiguously distinguish what is meant
by long and short distances.

\section{Quadratic expansion of the action}
\label{secqe}

We consider the action
\bea
S=Z \int d^3x \Bigg\{ \sqrt{- g} \Big[ \s R - 2 \Lambda_0 + \a R^2
+\b R_{\mu\nu}^2 \Big],
\label{action}
\eea
where $Z=1/(16\pi G)$ is the three-dimensional gravitational constant,
$\a, \b, \mu$ and $\s (= \pm 1)$ are constants, and $\Lambda_0$ is a cosmological constant.

We consider the action up to second order around the background spacetime
\bea
g_{\mu\nu}= \bar g_{\mu\nu} + h_{\mu\nu},
\label{fluc}
\eea
where we keep the formulae for arbitrary dimension $D$ for the moment
with future application to higher dimensions in mind.
Now the background $\bar g_{\mu\nu}$ is chosen to be a maximally symmetric
spacetime with the curvatures
\bea
\bar R_{\mu\nu}= \frac{R}{D}\, \bar g_{\mu\nu}, \quad
\bar R_{\mu\nu\rho\la}=\frac{R}{D(D-1)} (\bar g_{\mu\rho}\bar g_{\nu\la}
-\bar g_{\mu\la}\bar g_{\nu\rho}),
\label{back}
\eea
where $R=\pm\frac{D(D-1)}{\l^2}$ with the $+$ sign for de Sitter and $-$ sign for
anti-de Sitter spaces, and $l$ is the radius.
We define
\bea
h \equiv \bar g^{\mu\nu} h_{\mu\nu}.
\eea
Here and in what follows, bar indicates that the quantity stands for the background,
the indices are raised and lowered by the background metric $\bar g$,
the covariant derivative $\nabla$ is constructed with the background metric,
and the contraction is also understood by that.

We parametrize the metric in general dimensions as
\bea
h_{\mu\nu} = h_{\mu\nu}^T+\nabla_\mu \xi_\nu +\nabla_\nu \xi_\mu
 + \nabla_\mu \nabla_\nu \eta -\frac1D \bar g_{\mu\nu}\Box \eta + \frac1D \bar g_{\mu\nu}h,
\label{parad}
\eea
with
\bea
\nabla^\la h_{\la\mu}^T=0, \qquad
\bar g^{\mu\nu} h_{\mu\nu}^T=0, \qquad
\nabla^\la \xi_\la=0.
\eea
A straightforward calculation then yields the quadratic terms~\cite{Ohta1}
\bea
\frac{{\cal L}_2}{Z} \hs{-2}&=&\hs{-2} \frac14 h_{\mu\nu}^T \Big(\Box
 - \frac{2}{D(D-1)} R \Big)\Big[ \b\Big(\Box-\frac{2}{D(D-1)} R \Big)
+\frac{2}{D} R (D \a+\b)+\s \Big] h^{T,\mu\nu} \nn
&& +\frac{(D-1)(D-2)}{4D^2} \left[ \hat\eta \Delta \Box \hat\eta
- 2 h \Delta \sqrt{\Box\Big(\Box+\frac{1}{D-1} R \Big)}\; \hat\eta
+ h \Delta \Big(\Box +\frac{1}{D-1} R \Big) h \right] \nn
&& \hs{20} + \frac{\Lambda}{2} \Big[(h_{\mu\nu}^T)^2-2 \hat\xi_\mu^2
 +\frac{D-1}{D} \hat\eta^2 - \frac{D-2}{2D} h^2 \Big],
\label{actiond}
\eea
where we have defined
\bea
&& \hat \eta \equiv \sqrt{\Box\Big(\Box+\frac{1}{D-1} R\Big)}\; \eta, \quad
\hat\xi_\mu \equiv \sqrt{\Box+\frac{1}{D} R}\; \xi_\mu, \nn
&& \Delta \equiv \frac{4(D-1)\a+D\b}{D-2}\Box - \frac{2(D-4)}{D(D-2)}(D\a+\b)R -\s,\nn
&& \Lambda \equiv \Lambda_0-\frac{D-2}{2D} R\s-\frac{(D-4)}{2D^2}(D\a+\b)R^2.
\eea
These field redefinitions cancel the Jacobian introduced in the path integral
by the parametrization~\p{parad}.

Our next task is to introduce the gauge fixing and the corresponding FP ghost terms.
The BRST transformation for the fields is found to be
\bea
\d_B g_{\mu\nu} &=& -\d \la [ g_{\rho\nu}\nabla_\mu c^\rho + g_{\rho\mu}\nabla_\nu c^\rho
 + \nabla_\rho g_{\mu\nu} c^\rho], \nn
\d_B c^\mu &=& -\d\la c^\rho \nabla_\rho c^\mu, \nn
\d_B \bar c_\mu &=& i \d\la\, B_\mu,\nn
\d_B B_\mu &=& 0,
\label{brst}
\eea
which is nilpotent. Here $\d\la$ is an anticommuting parameter.
The gauge fixing term and the FP ghost terms are concisely written as
\bea
{\cal L}_{GF+FP}/Z \hs{-2}&=&\hs{-2} i \d_B [\bar c_\mu (\Box+c)
(\chi^\mu-\frac{a}{2} B^\mu)]/\d\la \nn
\hs{-2} &=&\hs{-2} - B_\mu (\Box+c) \chi^\mu
- i \bar c_\mu (\Box+c) \Big(\d_\mu^\nu \Box+\frac{1-b}{2}\nabla_\mu\nabla^\nu+R_\mu^\nu
\Big)c^\nu +\frac{a}{2} B_\mu (\Box+c) B^\mu,~~~~~~~~
\label{gfgh}
\eea
where $c$ is a constant, the indices are raised and lowered with the background metric, and
\bea
\chi_\nu = \nabla_\mu h^\mu{}_\nu-\frac{b+1}{4}\nabla_\nu h,
\eea
with $a$ and $b$ being gauge parameters.
The unusual factor $(\Box+c)$ is introduced in \p{gfgh} in order to
be able to diagonalize the kinetic terms of the gravitational fields
[see Eq.~\p{gaugechoice} below].
Eliminating the auxiliary fields $B_\mu$, we find the quadratic terms in the gauge fixing
are
\bea
{\cal L}_{GF} &=& -\frac{Z}{2a} \Big[
\hat\xi_\mu(\Box+c)\Big(\Box+ \frac{R}{D} \Big) \hat\xi^\mu
 -\Big(\frac{D-1}{D}\Big)^2 \hat\eta \Big(\Box+ \frac{R}{D-1} \Big)
\Big(\Box+ c + \frac{R}{D} \Big) \hat \eta \nn
&& - \frac{(D-1)(4-D-bD)}{2D^2} \hat\eta \sqrt{\Box\Big(\Box+ \frac{R}{D-1} \Big)}
 \Big(\Box+ c+\frac{R}{D} \Big) h \nn
&& - \frac{(4-D-bD)^2}{16 D^2} h \Box \Big(\Box+ c+\frac{R}{D} \Big) h
\Big].
\eea
Similarly we find the corresponding FP terms are given by
\bea
{\cal L}_{FP} = -i \Big[ \bar V^\mu (\Box+ c)\Big(\Box+ \frac{R}{D} \Big)V_\mu
-\hat{\bar S}\Big(\Box+ c+\frac{R}{D} \Big)
\Big(\frac{3-b}{2}\Box+ \frac{2}{D} R\Big) \hat S\Big],
\label{FP1}
\eea
where a factor $Z$ is dropped for simplicity (we can think that it is absorbed
into the fields) and we have defined
\bea
C_\mu \equiv V_\mu +\nabla_\mu S, \quad
\hat S = \sqrt{\Box}\; S, \quad
\nabla_\mu V^\mu=0, \nn
\bar C_\mu \equiv \bar V_\mu +\nabla_\mu {\bar S}, \quad
\hat{\bar S} = \sqrt{\Box}\; \bar S, \quad
\nabla_\mu \bar V^\mu=0.
\eea

Now we specialize to three dimensions. The quadratic and gauge fixing part of the action
is given by
\bea
{\cal L}_{2+GF}/Z\!\!\! &=&\!\!\! \frac14 h_{\mu\nu}^T \Big(\Box-\frac{R}{3}\Big)
\Big(\b\Box +\frac{6\a+\b}{3} R +\s \Big) h^{T\mu\nu} \nn
&& \hs{-5}
-\frac{1}{2a} \hat \xi_\mu(\Box+ c)\Big(\Box+\frac{R}{3}\Big) \hat \xi^\mu
+ \frac{1}{18} h\Big[\Delta\Big(\Box+\frac{R}{2}\Big)
+\frac{(1-3b)^2}{16a}\Box \Big\{\Box+ c +\frac{R}{3} \Big\}
\Big] h \nn
&& \hs{-5}
-\frac{1}{9} h \sqrt{\Box(\Box+R/2)}\Big[\Delta-\frac{1-3b}{2a}
\Big\{\Box + c+\frac{R}{3} \Big\}\Big] \hat\eta \nn
&& \hs{-5}
+\frac{1}{18} \hat\eta \Big[\Delta\Box+\frac{4}{a} \Big(\Box+\frac{R}{2} \Big)
\Big(\Box+ c +\frac{R}{3} \Big)\Big] \hat\eta
+\frac{\Lambda}{2} \Big[ (h_{\mu\nu}^T)^2-2 \hat\xi_\mu^2+\frac{2}{3}\hat \eta^2
-\frac{1}{6} h^2 \Big],~~~
\eea
where
\bea
\Delta &\equiv& (8\a+3\b)\Box + \frac{2(3\a+\b)}{3} R -\s, \nn
\Lambda &=& \Lambda_0 -\frac{\s R}{6}+\frac{3\a+\b}{18}R^2.
\label{para1}
\eea
Next we choose the parameters $a$ and $c$ such that the action is diagonalized:
\bea
1-3b = 2(8\a+3\b)a, ~~~
c+\frac{R}{3} = \frac{2(3\a+\b)R/3-\s}{8\a+3\b}.
\label{gaugechoice}
\eea
We keep the gauge parameter $b$ arbitrary in order to check the gauge dependence
of the result. Then the action simplifies to
\bea
{\cal L}_{2+GF}/Z\!\!\!\! &=&\!\!\!\! \frac14 h_{\mu\nu}^T \Delta_T h^{T\mu\nu}
-\frac{1}{(1-3b)} \hat \xi_\mu \Delta_\xi \hat \xi^\mu
+\frac{4}{3(1-3b)} \hat\eta \Delta_\eta \hat \eta
+ \frac{1}{6} h \Delta_h h ,
\label{diag}
\eea
where we have defined
\bea
\Delta_T &=& \Big(\Box-\frac{R}{3}\Big)
\Big( \b\Box +\frac{6\a+\b}{3}R +\s \Big) + 2 \Lambda , \nn
\Delta_\xi &=& \Big(\Delta-\frac{8\a+3\b}{3}R \Big)
\Big(\Box+\frac{R}{3}\Big) + (1-3b) \Lambda , \nn
\Delta_\eta &=& \Delta \Big(\frac{3-b}{8}\Box+\frac{R}{6}\Big)
+\frac{1-3b}{4}\Lambda ,\nn
\Delta_h &=& \Delta\Big(\frac{3-b}{8}\Box+\frac{R}{6}\Big)-\frac{1}{2}\Lambda.
\label{kin1}
\eea
It appears that the choice $b=3$ makes the kinetic operators simple.
However, in this gauge the uniform structure of the kinetic terms being quadratic
in the D'Alembertian is lost, and the method we use does not seem to be suitable
to deal with this case. Therefore in the rest of this paper we assume that $b<3$.

The FP terms~\p{FP1} become
\bea
{\cal L} = -i(\bar V^\mu \Delta_V V_\mu - \hat{\bar S} \Delta_S \hat S)
\eea
where
\bea
\Delta_V = \Big(\Box+\frac{2(3\a+\b)R/3 -\s}{8\a+3\b}-\frac{R}{3}\Big)
\Big(\Box+\frac{R}{3} \Big), \qquad
\Delta_S = \Delta \Big(\frac{3-b}{2}\Box+\frac{2R}{3}\Big).
\label{kin2}
\eea

We are now ready to evaluate the beta functions.

\section{Evaluation of the functional traces}
\label{seceval}

In this section, we set up the calculation of the RG equations for the space of
$S^3$ viewed as Euclideanized de Sitter space.

The quartic structure in the derivatives as opposed to quadratic in \cite{PS}
can be factorized into the product of two quadratic derivatives.
Thus we write the kinetic operators in a product forms of quadratic derivatives
\bea
\frac{1}{\b} \Delta_T = \Delta_{T1} \Delta_{T2} \equiv ( \Box-B_1^T)(\Box-B_2^T),
\label{factor}
\eea
and similarly for $\Delta_\xi, \Delta_\eta, \Delta_h, \Delta_V$ and $\Delta_S$,
where
\bea
B_{1,2}^T \hs{-2}&=&\hs{-2}
- \frac{ \s + 2R\a \pm \sqrt{[\s+2R (3\a+\b)/3]^2 -8\b\Lambda}}{2\b},\nn
B_{1,2}^\xi \hs{-2}&=&\hs{-2} \frac{ \s - 2R(3\a+\b)/3 \pm
\sqrt{[\s+2R(5\a+2\b)/3]^2 +4(3b-1)(8\a+3\b)\Lambda}}{2(8\a+3\b)},
\nn
B_{1,2}^\eta \hs{-2}&=&\hs{-2}  \frac{(b-3)\s -2[(3b-25)\a +(b-9)\b)] R/3}
{2(b-3)(8\a+3\b)} \nn
&& \pm
\frac{\sqrt{[(b-3)\s-2\{(3b+7)\a +(b+3)\b\} R/3]^2 +8(b-3)(3b-1)(8\a+3\b)\Lambda}}
{2(b-3)(8\a+3\b)},\nn
B_{1,2}^h \hs{-2}&=&\hs{-2}  \frac{ (b-3)\s - 2[(3b-25)\a+(b-9)\b)] R/3}
{2(b-3)(8\a+3\b)} \nn
&& \pm
\frac{\sqrt{[(b-3)\s-2\{(3b+7)\a+(b+3)\b\}R/3]^2 -16(b-3)(8\a+3\b)\Lambda}}
{2(b-3)(8\a+3\b)},
\nn
B_{1,2}^V \hs{-2}&=&\hs{-2} -\frac{R}{3},~~ \frac{\s+(2\a+\b)R/3}{8\a+3\b}, \nn
B_{1,2}^S \hs{-2}&=&\hs{-2} \frac{4R}{3(b-3)},~~ \frac{\s-2(3\a+\b)R/3}{8\a+3\b}.
\label{eigen1}
\eea

For each spin component, we choose the cutoff to be a function of the corresponding
operator given in \p{kin1}. The gauge fixed inverse propagator is
\bea
{\cal O} = Z \left( \begin{array}{cccc}
\frac14 \Delta_T & & & \\
 & -\frac{1}{1-3b} \Delta_\xi & & \\
 & & \frac{4}{3(1-3b)}\Delta_\eta & \\
 & & & \frac16 \Delta_h
\end{array}
\right).
\eea
The cutoff is chosen as
\bea
{\cal R}_k = Z \left( \begin{array}{cccc}
\frac14 {\cal R}_k(\Delta_T) & & & \\
 & -\frac{1}{1-3b} {\cal R}_k(\Delta_\xi) & & \\
 & & \frac{4}{3(1-3b)} {\cal R}_k(\Delta_\eta) & \\
 & & & \frac16 {\cal R}_k(\Delta_h)
\end{array}
\right).
\eea
We then have
\bea
{\cal O}+{\cal R}_k = Z \left( \begin{array}{cccc}
\frac14 P_k(\Delta_T) & & & \\
 & -\frac{1}{1-3b} P_k(\Delta_\xi) & & \\
 & & \frac{4}{3(1-3b)} P_k(\Delta_\eta) & \\
 & & & \frac16 P_k(\Delta_h)
\end{array}
\right),
\eea
where we have defined the function $P_k(z) = z+{\cal R}_k(z)$.
These formulae are valid for the case when the kinetic operators consist of
single D'Alembertian, but we should understand that they become a sum of those terms
when they are products of such operators.
We then find
\bea
\pa_t\G_k \hs{-2}&=&\hs{-2} \frac12 \Big[
\mbox{Tr} \{W(\Delta_{T1})+W(\Delta_{T2}) \}
+ \mbox{Tr} \{W(\Delta_{\xi 1})+W(\Delta_{\xi 2}) \}
+ \mbox{Tr} \{W(\Delta_{\eta 1})+W(\Delta_{\eta 2}) \} \nn
&& \hs{-8}  + \mbox{Tr} \{W(\Delta_{h1})+W(\Delta_{h2}) \} \Big]
- \Big[ \mbox{Tr} \{W(\Delta_{V1})+W(\Delta_{V2}) \}
+ \mbox{Tr} \{W(\Delta_{S1})+W(\Delta_{S2}) \} \Big],~~~~~~
\eea
where we have defined $W(z) = \frac{\pa_t {\cal R}_k}{P_k}$ and $t=\log k$.

Following \cite{Litim}, we use the optimized cutoff $R_k(z) = (k^2-|z|)\t(k^2-|z|)$.
We have $\pa_t R_k(z)=2k^2 \t(k^2-|z|)$, $P_k(z)= k^2$ for $z<k^2$
and $W(z)=2\t(k^2-|z|)$. Dividing the numerator and denominator by $k^2$, they are given by
\bea
\pa_t\G_k \hs{-2}&=&\hs{-2} \sum_n m_T \t(1-|\tilde\la_n^T|)
+ \sum_n m_\xi \t(1-|\tilde\la_n^\xi|)
+ \sum_n m_\eta \t(1-|\tilde\la_n^\eta|) \nn
&& + \sum_n m_h \t(1-|\tilde\la_n^h|)
-2 \sum_n m_V \t(1-|\tilde\la_n^V|)
-2 \sum_n m_S \t(1-|\tilde\la_n^S|),
\label{sum}
\eea
where $\tilde\la_n^{(i)} =\la_n^{(i)}/k^2$ are the distinct dimensionless eigenvalues of
the Euclideanized operator $\Delta_i$, and $m_n^{(i)}$ their multiplicities derived
in \cite{PS}.
The eigenvalues are given by
\bea
\la_{n1,2}^T \hs{-2}&=&\hs{-2} \frac{R}{6}(n^2+2n-2) +B_{1,2}^T,~~~~ n \geq 2, \nn
\la_{n1,2}^\xi \hs{-2}&=&\hs{-2} \frac{R}{6}(n^2+2n-1) + B_{1,2}^\xi,~~~~n \geq 2, \nn
\la_{n1,2}^\eta \hs{-2}&=&\hs{-2} \frac{R}{6}(n^2+2n) + B_{1,2}^\eta,~~~~n\geq 2 \nn
\la_{n1,2}^h \hs{-2}&=&\hs{-2} \frac{R}{6}(n^2+2n) + B_{1,2}^h,~~~~n\geq 0 \nn
\la_{n1,2}^V \hs{-2}&=&\hs{-2} \frac{R}{6}(n^2+2n-1) + B_{1,2}^V,~~~~n\geq 1 \nn
\la_{n1,2}^S \hs{-2}&=&\hs{-2} \frac{R}{6}(n^2+2n) + B_{1,2}^S,~~~~n\geq 1
\label{eigen2}
\eea
where the lower end of the allowed integer $n$ (denoted by $n_0$ below) for each
eigenvalue is also shown.
Here $B$'s are those defined in \p{eigen1}, but it should be understood that
we have rescaled the coupling constants by
\bea
G=\tilde G k^{-1},~~~
\Lambda_0 = \tL k^2,~~
\a= \ta k^{-2}, ~~
\b= \tb k^{-2},
\label{rescale}
\eea
so that they are dimensionless.
The multiplicities are
\bea
m_n^T \hs{-2}&=&\hs{-2} 2(n^2+2n-3), \nn
m_n^\xi \hs{-2}&=&\hs{-2}  m_n^V = 2(n^2+2n), \nn
m_n^\eta \hs{-2}&=&\hs{-2} m_n^h =m_n^S= (n+1)^2.
\eea

Each sum in \p{sum} is evaluated by the Euler-Maclaurin formula
\bea
\sum_{n=n_0}^\infty =\int_{n_0}^{n_{max}} F(n)dn + \frac12 F(n_0) -\frac{B_2}{2!} F'(n_0)
-\frac{B_4}{4!} F'''(n_0)+{\cal R},
\label{coeff}
\eea
where $B_n$ are the Bernoulli numbers ($B_2=1/6$ and $B_4=1/30$) and ${\cal R}$ is a remainder.
For the evaluation of the beta functions, only the first three terms are necessary.
Here $n_0$ are the lower ends of the allowed integer $n$ given in Eq.~\p{eigen2}
and $n_{max}$ are determined by
\bea
\tilde \la_n^{(i)} =1.
\eea

The evaluation of the sums is done using algebraic manipulation software.
As in Ref.~\cite{PS}, we restrict ourselves to the parameter region where $\Lambda$
and $R$ are of the same order. The reason is that the physics is contained in the
on-shell effective action though we have to be away from on-shell in order to
obtain the beta functions. We can be slightly off-shell for this purpose,
and then we can assume that $\Lambda$ and $R$ are of the same order.
Therefore we expand the coefficients in \p{coeff} in
powers of $\tL$ and keep at most terms linear in $\tL$ in $A$ while
we keep only the $\tL$-independent terms in $B$, and other terms of order $\tL R$
and $R^2$ are neglected.
Another reason why we do not look at the $R^2$ in this approach is
that even if we keep these terms, there is no way to tell which are terms
corresponding to $R^2$ and $R_{\mu\nu}^2$, so that we cannot obtain the beta
functions for $\a$ and $\b$ separately.
This is unavoidable as long as we evaluate the functional trace for the fixed
geometry of sphere.
In our approach, the evaluation of the functional trace for general background
would be difficult because the diagonalization of the quadratic terms of gauge-invariant
action and gauge fixing term is then more involved and cannot be done explicitly.

Restricting the terms this way makes the analysis of RG equations enormously simple.
In our approach we cannot see separately the running of the coupling constants $\a$ and $\b$
of the higher derivative terms, which correspond to terms of the order $k^{-1}R^2$.
Nevertheless we can check the coefficient of these terms.
A preliminary study shows that there are fixed points for these parameters;
we have checked that there are real solutions for $\ta$ and $\tb$ to the equations
setting the coefficient to zero. Though this is not sufficient for the existence
of the fixed point for $\ta$ and $\tb$, it is a necessary condition.
So here we content ourselves with assuming that they are fixed at a fixed point
in the following discussions.
We leave more detailed study of these terms for future work.

The result can be written as
\bea
\pa_t \G_k = \frac{V(S^3)}{16\pi}[ k^3 A + kB R + O(\tL R)],
\label{rgaction}
\eea
where $V(S^3)= 2\pi^2(\frac{6}{R})^{3/2}$ is the volume of $S^3$.
The coefficients are found to be
\bea
A &=& \frac{8}{\pi} \Bigg[ \frac{2}{3} \Big\{\Big(1 + \frac{\s}{\tb}\Big)^{\frac32}
-\Big(1-\frac{\s}{8\ta+ 3\tb}\Big)^{\frac32} \Big\}
+\frac{\s\tL}{b-3} \Big\{3b^2-5b-6 \nn
&& \hs{20} -2(b-3)\sqrt{1+\frac{\s}{\tb}} -(3b-7)b \sqrt{1-\frac{\s}{8\ta+3\tb}}\Bigg\}\Bigg]
\equiv A_0 + A_1\tL, \nn
B &=& \frac{4}{3\pi(b-3)(8\ta+3\tb)\tb} \Big[ (b-3) (96\ta^2+28\ta\tb-3\tb^2)
\sqrt{1+\frac{\s}{\tb}} +\{12(2b^2-5b+1)\ta \nn
&& \hs{10} +(9b^2-22b+3)\tb \}\tb \sqrt{1-\frac{\s}{8\ta+3\tb}}
+(36-5b-3b^2)(8\ta+3\tb)\tb \Big],
\label{ab}
\eea
Note that there is no term independent of $R$. This is to be expected
for the following reason. The result should contain only integer powers of $R$
and in a three-dimensional manifold without boundary the volume prefactor is
proportional to $R^{-3/2}$. This implies that the expansion of $\pa_t \G_k$
contains no $R$-independent term. This gives the first nontrivial check of our result.
There is a term of order $R^2$ with negative power of $k$.
This is because, as has been noted in \cite{Ohta2}, there is no divergence to
the $R^2$ and $R_{\mu\nu}^2$ terms in this renormalizable theory.
The coefficient of these terms is very complicated.

Evaluating the Euclidean version of the renormalized action~\p{action} on the $S^3$
background, we get
\bea
\G_k = V(S^3) \Big( \frac{2\Lambda_0}{16\pi G} -\frac{\s}{16\pi G}R
- \frac{3\a+\b}{48\pi G}R^2 \Big).
\label{eaction}
\eea
Applying the rescaling~\p{rescale} and comparing \p{rgaction} with \p{eaction}, we obtain
\bea
&& \frac{1}{8\pi} \Big( \frac{\pa_t \tL}{\tilde G} -\frac{\tL \pa_t \tilde G}{\tilde G^2}
+ \frac{3\tL}{\tilde G} \Big) = \frac{A}{16\pi}, \nn
&& \frac{\s}{16\pi} \Big( \frac{\pa_t \tilde G}{\tilde G^2} -\frac{1}{\tilde G} \Big)
 = \frac{B}{16\pi}.
\eea
Hence we find
\bea
\pa_t \tilde G = \tilde G(1+\s B \tilde G),~~
\pa_t \tL = -2\tL + \frac12 (A+2\s B \tL) \tilde G.
\label{rge}
\eea
Because we treat the coupling $\a$ and $\b$ as fixed, the RG equations
are considerably simplified. These equations are exactly of the same form
as in pure gravity with cosmological constant and in the topological gravity~\cite{PS}.

As a check, we examine if the running of the dimensionless combination
\bea
\pa_t(\tL \tilde G^2) =\frac12 (A+ 6\s B \tL) \tilde G^3,
\label{dimlessc}
\eea
is gauge independent. We find from \p{ab} that indeed this is independent of the gauge
parameter $b$.
\bea
\pa_t(\tL \tilde G^2)\hs{-2} &=&\hs{-2}
 \frac{8}{3\pi} \left[ \Big(1 + \frac{\s}{\tb}\Big)^{\frac32}
- \Big(1-\frac{\s}{8\ta+3\tb}\Big)^{\frac32}\right] \nn
&& \hs{-2} +\frac{4}{\pi}\s\left[
-10+3\frac{32\ta^2+4\ta\tb-3\tb^2}{(8\ta+3\tb)\tb} \sqrt{1 + \frac{\s}{\tb}}
-\frac{4\ta+\tb}{8\ta+3\tb} \sqrt{1 - \frac{\s}{8\ta+3\tb}} \right]\tL.~~~~
\eea
This gives the second nontrivial check of our results.

\section{Renormalization group flow and fixed points}
\label{secfix}

We now discuss the properties of the RG equations and
fixed points for the theory with $\s=\pm 1$ separately.

\subsection{$\s=+1$}

Let us first consider the case when the Einstein term has the usual sign.
Our RG equations~\p{rge} are well defined for the range
$
(\tb<-1~ \mbox{or}~ \tb>0),~
\mbox{and}~
(8\ta+3\tb<0~ \mbox{or}~ 8\ta+3\tb>1).
$
We shall be interested in the regions where unitary theories exist;
$(\ta \approx -\frac38 \tb, \tb>0)$ or $(\ta>0, \tb\approx 0)$.
So the most interesting region is $(\ta<0, \tb>0)$ and $(\ta>0,\tb>0)$.

\begin{figure}[htb]
\begin{center}
\begin{minipage}{60mm}
\includegraphics[width=6cm]{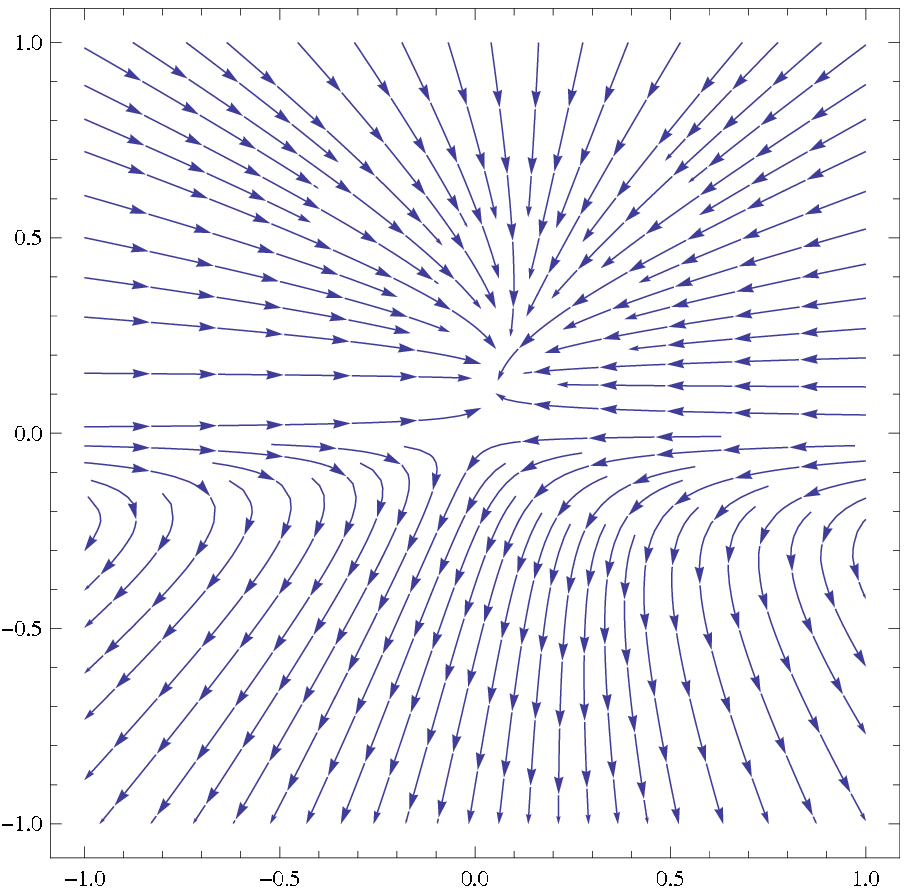}
\put(-85,170){$\tilde G$}
\put(5,80){$\tL$}
\put(-90,-15){(a)}
\end{minipage}
\hs{20}
\begin{minipage}{60mm}
\includegraphics[width=6cm]{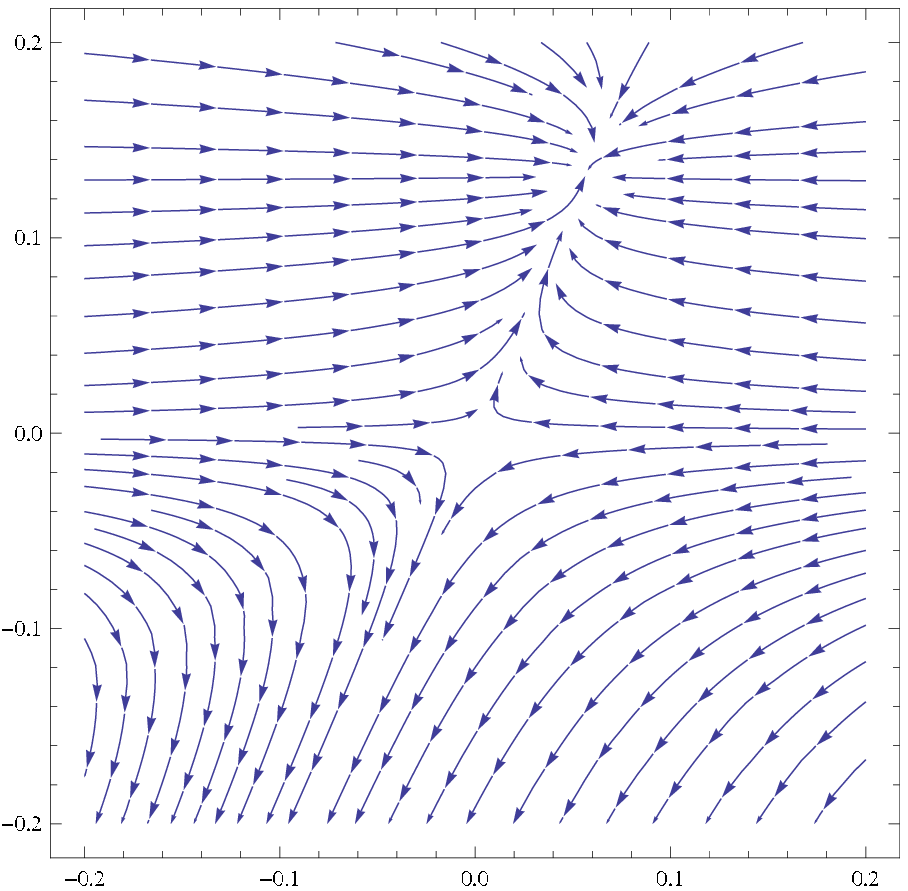}
\put(-85,170){$\tilde G$}
\put(5,80){$\tL$}
\put(-90,-15){(b)}
\end{minipage}
\caption{
The RG flow (a) for $\ta=-1,\tb=3$ and $b=0$.
(b) Magnified one near origin.
}
\label{fig1-1}
\end{center}
\end{figure}

The RE equations have two fixed points. One is the Gaussian
fixed point $\tilde G = \tL=0$, which is seen to be attractive in the $\tL$
direction and repulsive in the $\tilde G$ direction.
The other fixed point is at
\bea
\tilde G_* = -\frac{1}{B}, ~~
\tL_* = \frac{A_0}{A_1+6B},
\eea
where the constants $A_0, A_1$ and $B$ are defined in Eq.~\p{ab}.
The constants $G$ and $\Lambda$ are dimensionful parameters,
and on general ground we expect the fixed points depend on the gauge.
To get some idea what values they typically have, we give their values
$\tilde G_* \sim 0.133$ and $\tL_* \sim 0.057$ for $\ta=-1, \tb=3, b=0$.
The flow is shown in Fig.~\ref{fig1-1} for this choice of $\ta,\tb$ and $b$.

\begin{figure}[hbt]
\begin{center}
\begin{minipage}{70mm}
\includegraphics[width=5cm]{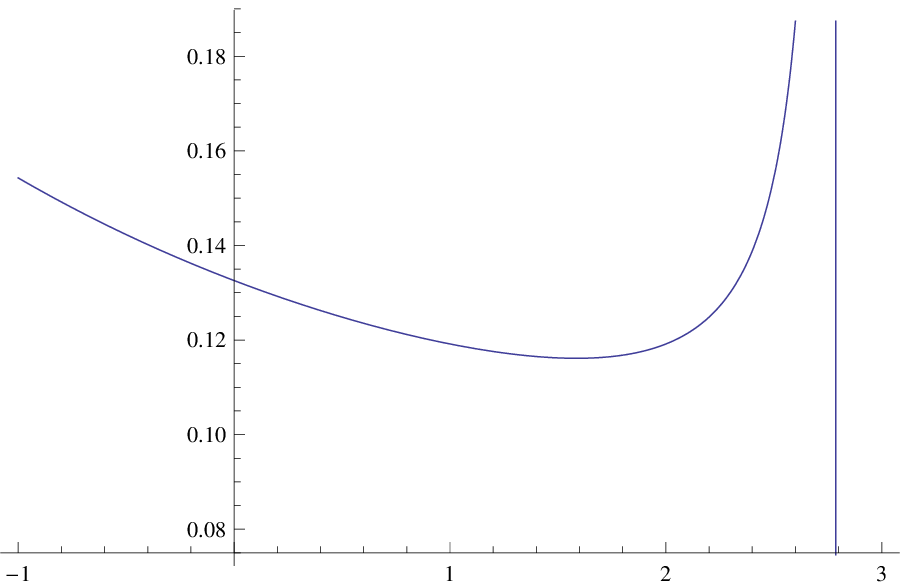}
\put(-130,120){$\tilde G_*$}
\put(5,5){$b$}
\put(-75,-10){(a)}
\end{minipage}
\hs{10}
\begin{minipage}{60mm}
\includegraphics[width=6cm]{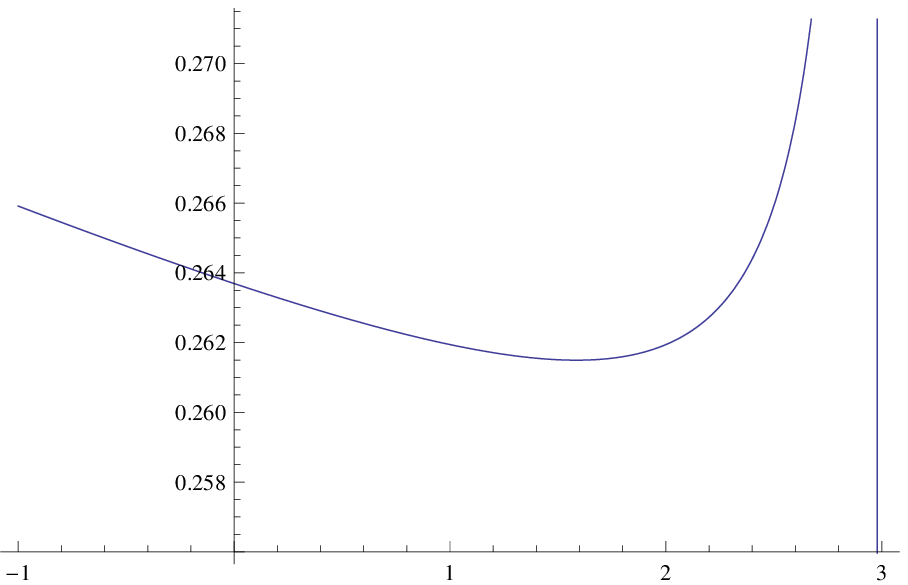}
\put(-130,120){$\tilde G_*$}
\put(5,5){$b$}
\put(-55,-10){(b)}
\end{minipage}
\caption{
Gauge dependence of the nontrivial fixed point of $\tilde G_*$
(a) for $\ta=-1,\tb=3$ and (b) for $\ta=1,\tb=3$.
}
\label{fig1-2}
\end{center}
\end{figure}

We have checked how much these fixed points are dependent on the gauge parameter $b$.
We show how $\tilde G_*$ changes for $-1<b<3$ in Fig.~\ref{fig1-2} (a)
for $\ta=-1, \tb=3$ and in (b) for $\ta=1, \tb=3$.
These parameters are chosen because we consider that it is natural to choose
these of the order 1.
We see that they both give qualitatively similar values and behaviors.
In particular the absolute value is very small well below $b\sim 3$, so that this is
certainly within the perturbative domain. Therefore, though our result is based on
a one-loop calculation, we can expect that it correctly represents the features of
the Wilsonian RG flow.

\begin{figure}[hbt]
\begin{center}
\begin{minipage}{60mm}
\includegraphics[width=6cm]{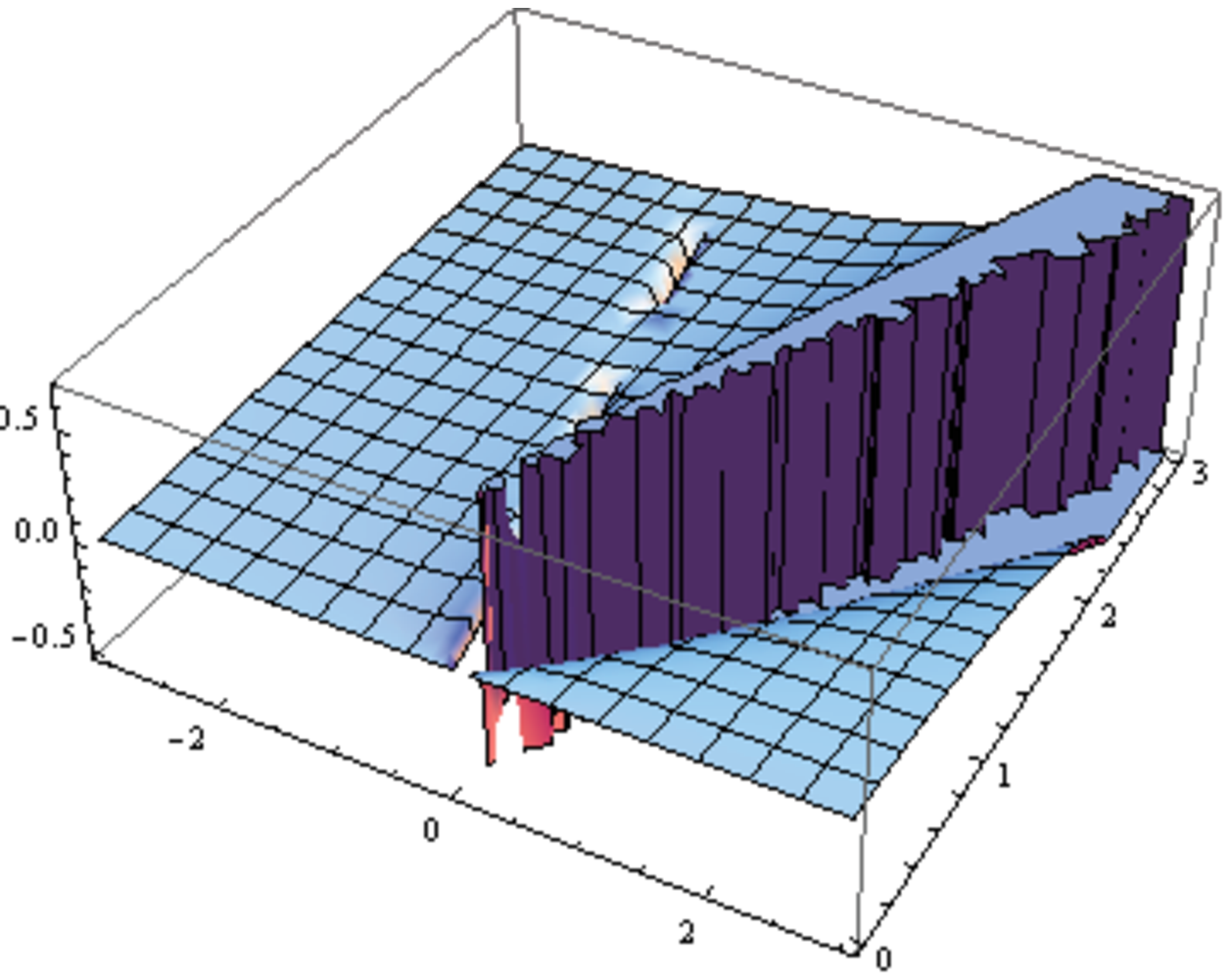}
\put(-170,95){$\tilde G$}
\put(-120,10){$\ta$}
\put(-10,30){$\tb$}
\put(-90,-20){(a)}
\end{minipage}
\hs{20}
\begin{minipage}{60mm}
\includegraphics[width=6cm]{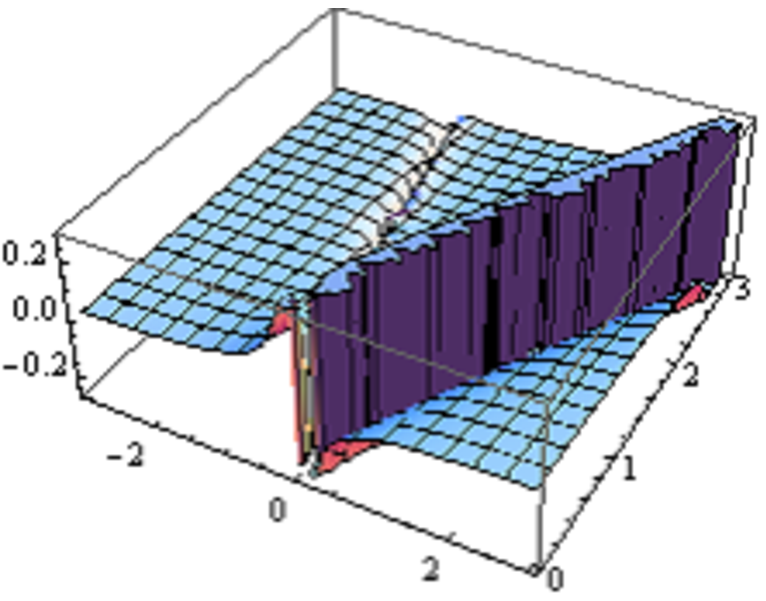}
\put(-170,95){$\tL$}
\put(-120,10){$\ta$}
\put(-10,30){$\tb$}
\put(-90,-20){(b)}
\end{minipage}
\caption{
The fixed point values of (a) $\tilde G_*$ and (b) $\tL_*$ for $b=0$
as a function of $\ta$ and $\tb$.
}
\label{fig1-3}
\end{center}
\end{figure}

It is remarkable that the $b$-dependence disappears from the fixed point
of the cosmological constant.
As we saw in the preceding section, only $A_1$ and $B$ depend on $b$,
but the combination $A_1+6B$ is precisely that appearing in the beta function of
the dimensionless coupling~\p{dimlessc}, and hence that dependence drops out.
We find that for the typical choice of $\ta$ and $\tb$, the cosmological constant
is positive and small; $0.057$ for $\ta=-1$ and $\tb=3$, and 0.045 for $\ta=1$ and $\tb=3$.
The precise values may not be so important, but the fact that they are generally
rather small may have some significance.
Fig.~\ref{fig1-3} shows how the values of $\tilde G$ and $\tL$ change for
$0<\ta<3$ and $0<\tb<3$.
This figure shows that the values are indeed small for a wide range of $\ta$ and $\tb$
except near $8\ta+3\tb=0$ and $\tb=0$, where these quantities are singular.

\subsection{$\s=-1$}

Let us now turn to the opposite sign for the Einstein term.
Our RG equations~\p{rge} are well defined for the range
$
(\tb<0~ \mbox{or}~ \tb>1),~
\mbox{and}~
(8\ta+3\tb<-1~ \mbox{or}~ 8\ta+3\tb>0).
$
We shall be interested in the regions where unitary theories exist;
$(\ta \approx -\frac38 \tb, \tb>0)$ or $(\ta>0, \tb\approx 0)$.
So the most interesting region is $(\ta<0, \tb>1)$ and $(\ta>0,\tb>0)$.

\begin{figure}[tb]
\begin{center}
\begin{minipage}{60mm}
\includegraphics[width=6cm]{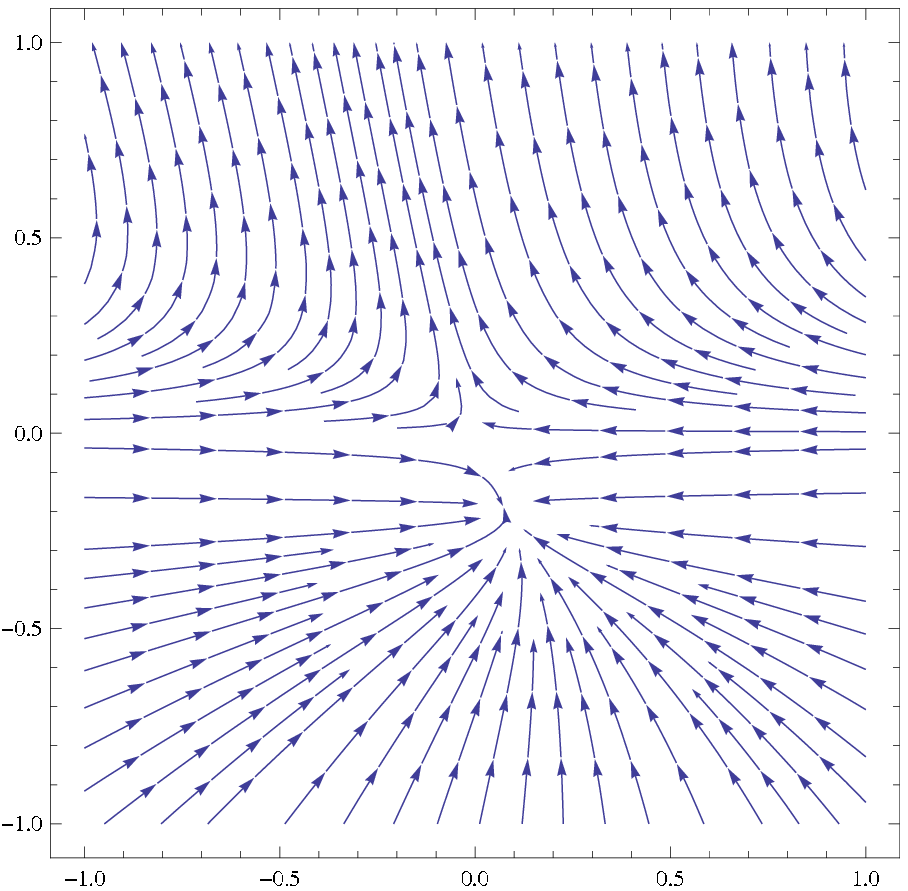}
\put(-85,170){$\tilde G$}
\put(5,80){$\tL$}
\put(-90,-15){(a)}
\end{minipage}
\hs{10}
\begin{minipage}{60mm}
\includegraphics[width=6cm]{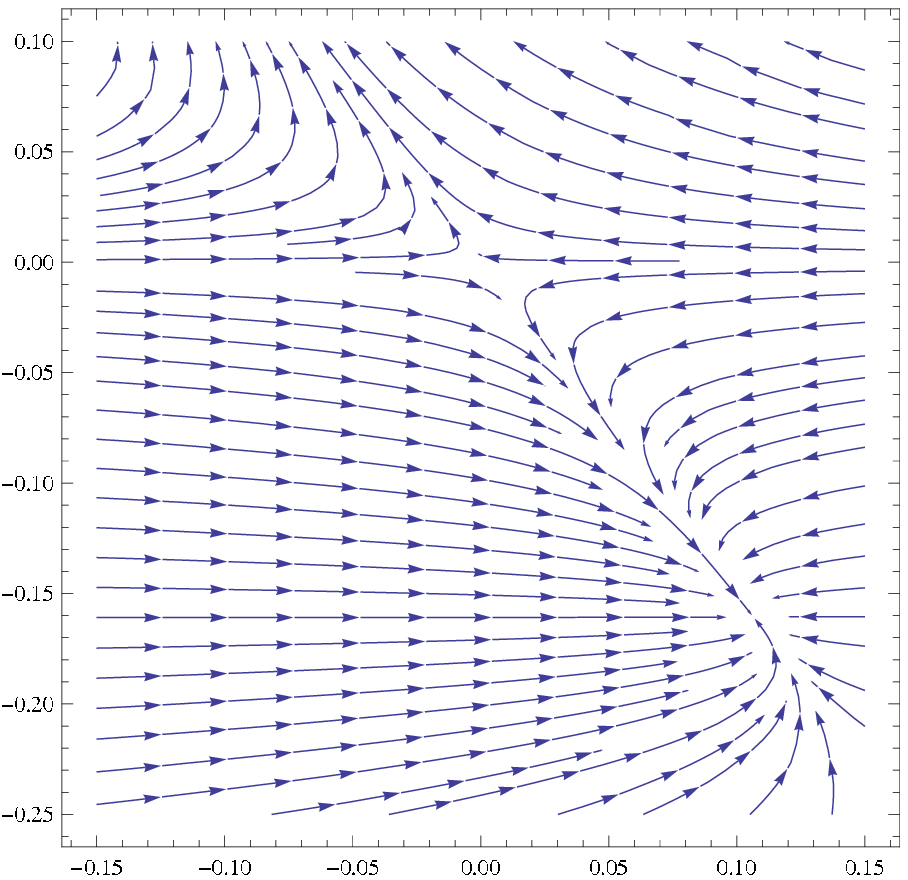}
\put(-85,170){$\tilde G$}
\put(5,80){$\tL$}
\put(-90,-15){(a)}
\end{minipage}
\caption{
The flow is shown for $\ta=-1,\tb=3$ and $b=0$. (b) is a magnified diagram.}
\label{figm-1}
\end{center}
\end{figure}

\begin{figure}[tb]
\begin{center}
\begin{minipage}{60mm}
\includegraphics[width=5cm]{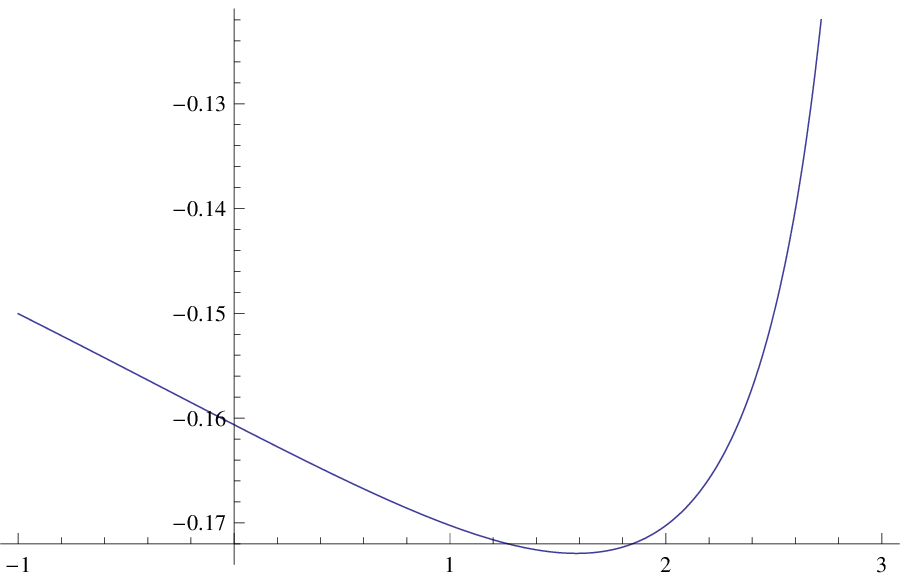}
\put(-110,100){$\tilde G_*$}
\put(5,5){$b$}
\put(-75,-10){(a)}
\end{minipage}
\hspace*{10mm}
\begin{minipage}{60mm}
\includegraphics[width=5cm]{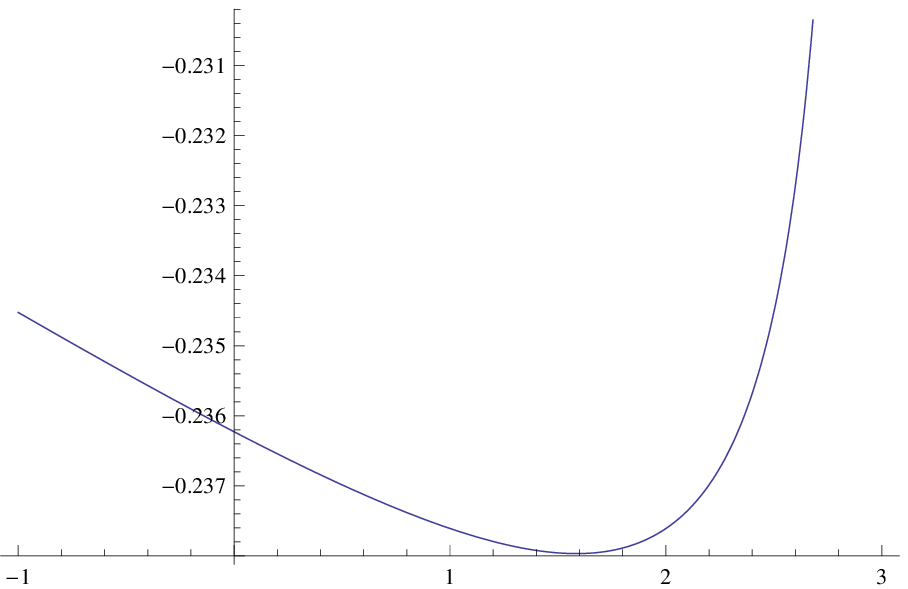}
\put(-110,100){$\tilde G_*$}
\put(5,5){$b$}
\put(-55,-10){(b)}
\end{minipage}
\caption{Gauge dependence of the concrete values of the nontrivial fixed point of $\tilde G_*$
(a) for $\ta=-1,\tb=3$ and (b) for $\ta=1,\tb=3$.}
\label{figm-2}
\end{center}
\end{figure}

Here again there is the Gaussian fixed point which is attractive in the $\tL$
direction and repulsive in the $\tilde G$ direction.
Another nontrivial fixed point in this case is given by
\bea
\tilde G_* = \frac{1}{B}, ~~
\tL_* = \frac{A_0}{-A_1+6B}.
\eea
Let us see again the concrete values and the gauge dependence of the fixed point
of the gravitational constant.
For $\ta=-1, \tb=3, b=0$, we find $\tilde G_* \sim -0.161$ and $\tL_* \sim 0.106$.
The flow is shown in Fig.~\ref{figm-1}.
We show how $\tilde G_*$ changes for $-1<b<3$ in Fig.~\ref{figm-2} (a)
for $\ta=-1, \tb=3$ and in (b) for $\ta=1, \tb=3$.
We see that again they both give qualitatively similar values and behaviors.
In particular the absolute value is again very small.

We note that the sign of the Newton constant changes from $\s=+1$ case and
it takes negative value for typical values of $\ta$ and $\tb$.
Recall that in this case we took the negative sign for the Einstein term,
but the fixed point of the gravitational constant takes negative values,
resulting in positive Einstein term.
Namely the Einstein term has positive coefficient at the fixed point
even if we start with negative sign.

\begin{figure}[h]
\begin{center}
\begin{minipage}{60mm}
\includegraphics[width=6cm]{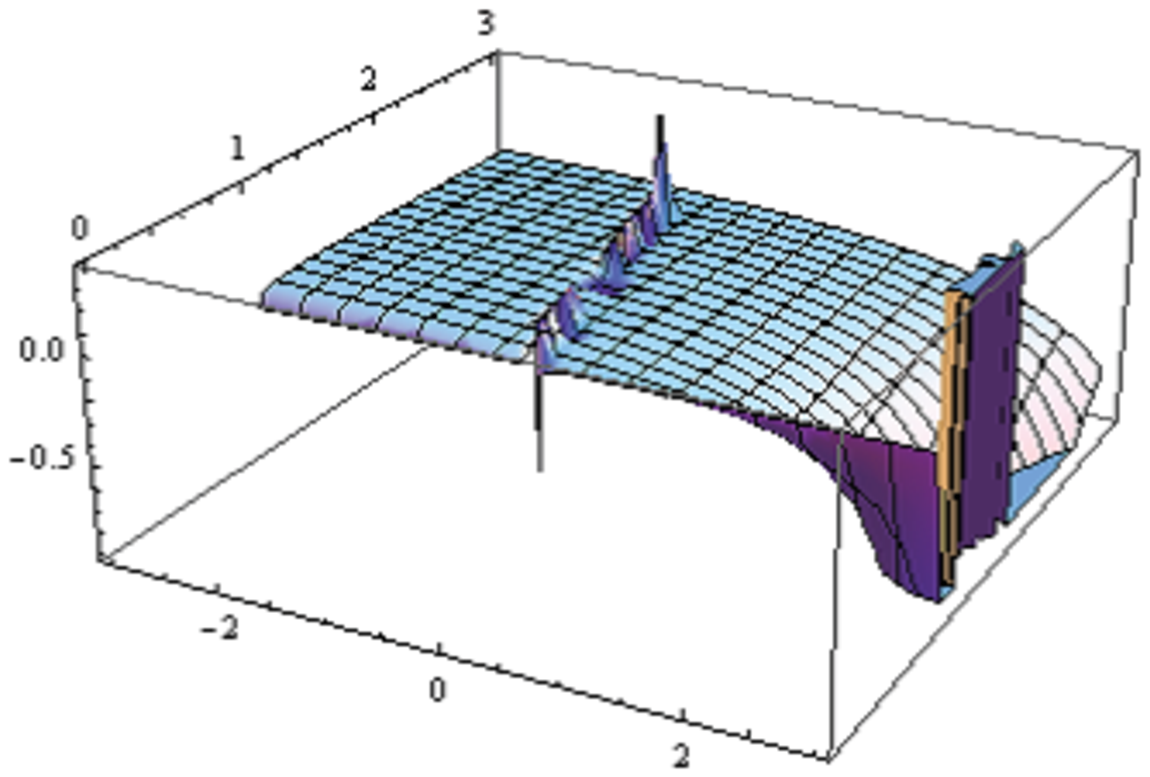}
\put(-170,95){$\tilde G_*$}
\put(-120,0){$\ta$}
\put(-20,20){$\tb$}
\put(-90,-20){(a)}
\end{minipage}
\hs{20}
\begin{minipage}{60mm}
\includegraphics[width=6cm]{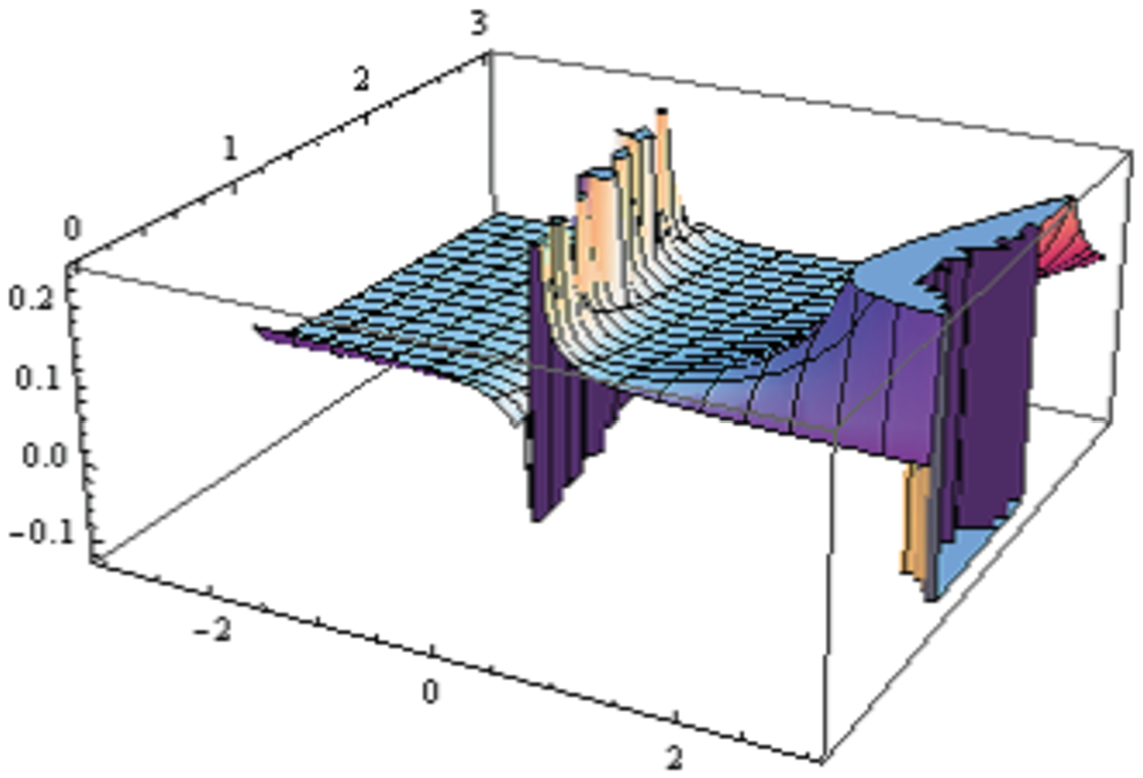}
\put(-170,95){$\tL$}
\put(-120,0){$\ta$}
\put(-20,20){$\tb$}
\put(-90,-20){(b)}
\end{minipage}
\caption{
The fixed point values of (a) $\tilde G_*$ and (b) $\tL_*$ for $b=0$
as a function of $\ta$ and $\tb$.
}
\label{figm-3}
\end{center}
\end{figure}

It is remarkable that the fixed point of the cosmological constant
again does not depend on the gauge. We find that its values are
$0.106$ for $\ta=-1,\tb=3$ and $0.038$ for  $\ta=1,\tb=3$,
again very small positive numbers.
Fig.~\ref{figm-3} shows how the values of $\tilde G$ and $\tL$ changes for
$-3<\ta<3$ and $0<\tb<3$.

\section{Discussions and conclusions}
\label{secconc}

In this paper we have studied the quantum effects of three-dimensional higher derivative
gravity which has attracted much attention recently. In particular we have
obtained the Wilsonian RG equations to study their fixed points.
Though we did not attempt to derive those for the coefficients of the higher curvature
terms because we cannot derive those for $\ta$ and $\tb$ separately and also
they become awfully complicated, we have some evidence that there is some fixed point
for $\ta$ and $\tb$. There are several conclusions we can draw even within this restriction.
Assuming that $\ta$ and $\tb$ have fixed points, we have found that there are ultraviolet
fixed points for the gravitational and cosmological constants, one the usual Gaussian
and the other nontrivial one for both signs of the Einstein term.
This shows that this theory is asymptotically safe.

We have also found that the fixed point value of the gravitational constant is generically
small, so that our one-loop discussion may well be justified.
What is interesting is that if we take the negative sign for the Einstein term
in the bare action, the gravitational constant has only a fixed point of negative values
or zero, in the former case resulting in positive Einstein term.
This is an interesting result which has not been explored in other approaches.
It would be extremely interesting whether this happens also in four-dimensional gravity.

We have also shown that RG equations are singular for the parameters,
$8\ta+3\tb=0$ or $\tb=0$, corresponding to the new massive gravity or a special case
of $f(R)$ gravity.
Indeed, we see from Figs.~\ref{fig1-3} and \ref{figm-3} that $\tilde G_*$ and $\tL_*$
diverge there.
If we look at the diagonalized action~\p{diag}, we find that
only the kinetic term for the transverse traceless mode is quadratic in the D'Alembertian
but the rest are all linear, as can be seen from \p{kin1}, and the limit is singular.
This means that the new massive gravity and $f(R)$ gravity would not correspond to
a fixed point in this class of theories.
We would like to note that this result is one of the conclusions
that are obtained within the approximation that the coefficients
of the higher curvature terms $\ta$ and $\tb$ are not subject to the flow, so strictly
speaking there may be a possibility that the picture might be different in a larger space
when these parameters are allowed to run.

Thus, though we did not attempt to evaluate the beta functions for $\ta$ and $\tb$,
it is quite interesting to examine whether the picture may change if we include
those parameters in the RG analysis. For this purpose, we have to find
a way to compute these for general backgrounds, not just for sphere,
because we have to tell which is the contribution to $R^2$ and which is to $R_{\mu\nu}^2$.
Probably the heat kernel method may be useful in such a study.
We hope to return to this problem in the future.

\section*{Acknowledgement}

We would like to thank Roberto Percacci for valuable discussions
and careful reading of the manuscript.
This work was supported in part by the Grant-in-Aid for
Scientific Research Fund of the JSPS (C) No. 24540290 and (A) No. 22244030.

\end{document}